\documentclass{svmult}

\usepackage{url}
\usepackage[numbers]{natbib}
\usepackage{amsmath}
\usepackage{amssymb}
\usepackage{eqname}
\usepackage{mathtools}
\usepackage{boxedminipage}

\begin{document}
\message{<Paul Taylor's Proof Trees, 2 August 1996>}

\def\introrule{{\cal I}}\def\elimrule{{\cal E}}
\def\andintro{\using{\land}\introrule\justifies}
\def\impelim{\using{\Rightarrow}\elimrule\justifies}
\def\allintro{\using{\forall}\introrule\justifies}
\def\allelim{\using{\forall}\elimrule\justifies}
\def\falseelim{\using{\bot}\elimrule\justifies}
\def\existsintro{\using{\exists}\introrule\justifies}

\def\andelim#1{\using{\land}#1\elimrule\justifies}
\def\orintro#1{\using{\lor}#1\introrule\justifies}

\def\impintro#1{\using{\Rightarrow}\introrule_{#1}\justifies}
\def\orelim#1{\using{\lor}\elimrule_{#1}\justifies}
\def\existselim#1{\using{\exists}\elimrule_{#1}\justifies}


\newdimen\proofrulebreadth \proofrulebreadth=.05em
\newdimen\proofdotseparation \proofdotseparation=1.25ex
\newdimen\proofrulebaseline \proofrulebaseline=2ex
\newcount\proofdotnumber \proofdotnumber=3
\let\then\relax
\def\hfi{\hskip0pt plus.0001fil}
\mathchardef\squigto="3A3B
%
\newif\ifinsideprooftree\insideprooftreefalse
\newif\ifonleftofproofrule\onleftofproofrulefalse
\newif\ifproofdots\proofdotsfalse
\newif\ifdoubleproof\doubleprooffalse
\let\wereinproofbit\relax
%
\newdimen\shortenproofleft
\newdimen\shortenproofright
\newdimen\proofbelowshift
\newbox\proofabove
\newbox\proofbelow
\newbox\proofrulename
%
\def\shiftproofbelow{\let\next\relax\afterassignment\setshiftproofbelow\dimen0 }
\def\shiftproofbelowneg{\def\next{\multiply\dimen0 by-1 }%
\afterassignment\setshiftproofbelow\dimen0 }
\def\setshiftproofbelow{\next\proofbelowshift=\dimen0 }
\def\setproofrulebreadth{\proofrulebreadth}

\def\prooftree{
%
\ifnum  \lastpenalty=1
\then   \unpenalty
\else   \onleftofproofrulefalse
\fi
%
\ifonleftofproofrule
\else   \ifinsideprooftree
        \then   \hskip.5em plus1fil
        \fi
\fi
%
\bgroup
\setbox\proofbelow=\hbox{}\setbox\proofrulename=\hbox{}%
\let\justifies\proofover\let\leadsto\proofoverdots\let\Justifies\proofoverdbl
\let\using\proofusing\let\[\prooftree
\ifinsideprooftree\let\]\endprooftree\fi
\proofdotsfalse\doubleprooffalse
\let\thickness\setproofrulebreadth
\let\shiftright\shiftproofbelow \let\shift\shiftproofbelow
\let\shiftleft\shiftproofbelowneg
\let\ifwasinsideprooftree\ifinsideprooftree
\insideprooftreetrue
%
\setbox\proofabove=\hbox\bgroup$\displaystyle 
\let\wereinproofbit\prooftree
%
\shortenproofleft=0pt \shortenproofright=0pt \proofbelowshift=0pt
%
\onleftofproofruletrue\penalty1
}

\def\eproofbit{
%
\ifx    \wereinproofbit\prooftree
\then   \ifcase \lastpenalty
        \then   \shortenproofright=0pt  
        \or     \unpenalty\hfil         
        \or     \unpenalty\unskip       
        \else   \shortenproofright=0pt  
        \fi
\fi
%
\global\dimen0=\shortenproofleft
\global\dimen1=\shortenproofright
\global\dimen2=\proofrulebreadth
\global\dimen3=\proofbelowshift
\global\dimen4=\proofdotseparation
\global\count255=\proofdotnumber
%
$\egroup  
%
\shortenproofleft=\dimen0
\shortenproofright=\dimen1
\proofrulebreadth=\dimen2
\proofbelowshift=\dimen3
\proofdotseparation=\dimen4
\proofdotnumber=\count255
}

\def\proofover{
\eproofbit 
\setbox\proofbelow=\hbox\bgroup 
\let\wereinproofbit\proofover
$\displaystyle
}%
%
\def\proofoverdbl{
\eproofbit 
\doubleprooftrue
\setbox\proofbelow=\hbox\bgroup 
\let\wereinproofbit\proofoverdbl
$\displaystyle
}%
%
\def\proofoverdots{
\eproofbit 
\proofdotstrue
\setbox\proofbelow=\hbox\bgroup 
\let\wereinproofbit\proofoverdots
$\displaystyle
}%
%
\def\proofusing{
\eproofbit 
\setbox\proofrulename=\hbox\bgroup 
\let\wereinproofbit\proofusing
\kern0.3em$
}

\def\endprooftree{
\eproofbit 
  \dimen5 =0pt
%
\dimen0=\wd\proofabove \advance\dimen0-\shortenproofleft
\advance\dimen0-\shortenproofright
%
\dimen1=.5\dimen0 \advance\dimen1-.5\wd\proofbelow
\dimen4=\dimen1
\advance\dimen1\proofbelowshift \advance\dimen4-\proofbelowshift
%
\ifdim  \dimen1<0pt
\then   \advance\shortenproofleft\dimen1
        \advance\dimen0-\dimen1
        \dimen1=0pt
        \ifdim  \shortenproofleft<0pt
        \then   \setbox\proofabove=\hbox{%
                        \kern-\shortenproofleft\unhbox\proofabove}%
                \shortenproofleft=0pt
        \fi
\fi
%
\ifdim  \dimen4<0pt
\then   \advance\shortenproofright\dimen4
        \advance\dimen0-\dimen4
        \dimen4=0pt
\fi
%
\ifdim  \shortenproofright<\wd\proofrulename
\then   \shortenproofright=\wd\proofrulename
\fi
%
\dimen2=\shortenproofleft \advance\dimen2 by\dimen1
\dimen3=\shortenproofright\advance\dimen3 by\dimen4
%
\ifproofdots
\then
        \dimen6=\shortenproofleft \advance\dimen6 .5\dimen0
        \setbox1=\vbox to\proofdotseparation{\vss\hbox{$\cdot$}\vss}%
        \setbox0=\hbox{%
                \advance\dimen6-.5\wd1
                \kern\dimen6
                $\vcenter to\proofdotnumber\proofdotseparation
                        {\leaders\box1\vfill}$%
                \unhbox\proofrulename}%
\else   \dimen6=\fontdimen22\the\textfont2 
        \dimen7=\dimen6
        \advance\dimen6by.5\proofrulebreadth
        \advance\dimen7by-.5\proofrulebreadth
        \setbox0=\hbox{%
                \kern\shortenproofleft
                \ifdoubleproof
                \then   \hbox to\dimen0{%
                        $\mathsurround0pt\mathord=\mkern-6mu%
                        \cleaders\hbox{$\mkern-2mu=\mkern-2mu$}\hfill
                        \mkern-6mu\mathord=$}%
                \else   \vrule height\dimen6 depth-\dimen7 width\dimen0
                \fi
                \unhbox\proofrulename}%
        \ht0=\dimen6 \dp0=-\dimen7
\fi
%
\let\doll\relax
\ifwasinsideprooftree
\then   \let\VBOX\vbox
\else   \ifmmode\else$\let\doll=$\fi
        \let\VBOX\vcenter
\fi
\VBOX   {\baselineskip\proofrulebaseline \lineskip.2ex
        \expandafter\lineskiplimit\ifproofdots0ex\else-0.6ex\fi
        \hbox   spread\dimen5   {\hfi\unhbox\proofabove\hfi}%
        \hbox{\box0}%
        \hbox   {\kern\dimen2 \box\proofbelow}}\doll%
%
\global\dimen2=\dimen2
\global\dimen3=\dimen3
\egroup 
\ifonleftofproofrule
\then   \shortenproofleft=\dimen2
\fi
\shortenproofright=\dimen3
%
\onleftofproofrulefalse
\ifinsideprooftree
\then   \hskip.5em plus 1fil \penalty2
\fi
}


\title*{Nominalistic Logic (Extended Abstract)}

\author{J{\o}rgen Villadsen}

\institute{Department of Informatics and Mathematical Modeling \\ Technical University of Denmark} 

\maketitle

\pagestyle{plain}
\thispagestyle{plain}

\vspace{-2ex}

Nominalistic Logic (NL) is a new presentation of Paul Gilmore's Intensional Type Theory (ITT) as a sequent calculus
together with a succinct nominalization axiom (N) that permits names of predicates as individuals in certain cases.
The logic has a flexible comprehension axiom, but no extensionality axiom and no infinity axiom,
although axiom N is the key to the derivation of Peano's postulates for the natural numbers.

\medskip

\noindent References:

\smallskip

\noindent ITT \\[.25ex] Paul Gilmore:
\emph{An Intensional Type Theory: Motivation and Cut-Elimination.}
Journal of Symbolic Logic 383-400 2001.

\smallskip

\noindent NL \\[.25ex] J{\o}rgen Villadsen:
\emph{Nominalistic Logic: From Naive Set Theory to Intensional Type Theory.}
Pages 57--85 in Klaus Robering (editor): New Approaches to Classes and Concepts.
Volume 14, Studies in Logic, College Publications 2008.

\

\begin{center}
\begin{boxedminipage}{\textwidth}
\begin{center}
\begin{minipage}{.96\textwidth}
\vspace{2.5ex}
\begin{center}
$c$ and $x, y, z, \ldots, x_n, y_n, z_n, \ldots$ range over constants and countably infinitely \\
many variables, respectively, and $p, q, r, s, t, \ldots, p_n, q_n, r_n, s_n, t_n, \ldots$ \\
range over terms produced by the grammar:
\smallskip
\begin{eqnarray*}
t & \,\Coloneqq\, & pt \mid \lambda x. p \mid x \mid c
\end{eqnarray*}
\\[1ex]
$p t_1 \cdots t_n$ stands for $(p t_1) \cdots t_n$
\\[2ex]
$\lambda x_1 \cdots x_n. p$ stands for $\lambda x_1. \cdots \lambda
x_n. p$ (used for all variable-binding operators).
\\[2ex]
A variable $x$ occurs bound (free) in a term if $x$ is (not) in the
scope of a $\lambda x$.
\\[1ex]
The term $p[t/x]$ is $p$ with every free occurrence of $x$ replaced
with $t$.
\\[2ex]
$\alpha$-renaming of the variable $x$ to the variable $y$ in $\lambda x. p$ is $\lambda y. p[y/x]$ assuming $y$ \\
is not free in $p$ and does not become bound in $p$.
\\[1ex]
$\beta$-reduction of $(\lambda x. p)t$ is $p[t/x]$ assuming no free
variable in $t$ becomes bound.
\\[1ex]
$\eta$-reduction of $\lambda x. px$ is $p$ assuming $x$ is not free in
$p$.
\\[2ex]
$s \sim t$ if $t$ is $s$ with a series of $\alpha$-renamings.
\\[1ex]
$s \succ t$ if $t$ is $s$ with either a $\beta$-reduction or an
$\eta$-reduction.
\end{center}
\vspace{0.5ex}
\end{minipage}
\end{center}
\end{boxedminipage}
\end{center}

\begin{center}
\begin{boxedminipage}{\textwidth}
\begin{center}
\begin{minipage}{.96\textwidth}
\vspace{2.5ex}
\begin{center}
$\tau, \ldots, \tau_n, \ldots$ and $\sigma, \ldots, \sigma_n, \ldots$ range over types and predicate types, \\
respectively, produced by the grammar:
\smallskip
\begin{eqnarray*}
\tau & \,\Coloneqq\, & \sigma \mid \imath
\\[1ex]
\sigma & \,\Coloneqq\, & \tau \sigma \mid o
\end{eqnarray*}
\\[1ex]
$\tau_1 \cdots \tau_n \sigma$ stands for $\tau_1 \cdots (\tau_n \sigma)$
\\[2ex]
It is assumed that every constant and variable has a unique type \\
such that there are infinitely many variables for each type
\\[2ex]
The type system $t : \tau$ determines that a term $t$ has type $\tau$ by the conditions:
\begin{eqnarray}
& x : \tau & \text{if $x$ has type $\tau$}
\eqname{var}
\\[1ex]
& c : \tau & \text{if $c$ has type $\tau$}
\eqname{con}
\\[1ex]
p : \tau \sigma \, ~\text{and}~ \, t : \tau & ~\text{gives}~ & pt : \sigma
\eqname{app}
\\[1ex]
x : \tau \, ~\text{and}~ \, p : \sigma & ~\text{gives}~ & \lambda x. p : \tau \sigma
\eqname{abs}
\\[1ex]
& p : \imath & \text{if $p$ is nominalizable}
\eqname{nom}
\end{eqnarray}
\\[1ex]
$p$ is nominalizable if $p : \sigma$ and $x : \imath$ for every free variable $x$ in $p$
\\[2ex]
$p$ is a formula if $p : o$ (a sentence is a closed formula)
\end{center}
\vspace{0.5ex}
\end{minipage}
\end{center}
\end{boxedminipage}
\end{center}

\begin{center}
\begin{boxedminipage}{\textwidth}
\begin{center}
\begin{minipage}{.96\textwidth}
\vspace{2.5ex}
\begin{center}
There is a constant of type $ooo$ and for every type $\tau$ a constant of type $(\tau o)o$
\\[2ex]
The constants and variables must have appropriate types in the abbreviations:
\begin{displaymath}
\begin{array}{r@{~\,\coloneqq\,~}l@{\hspace{2em}}l}
  \neg p & cpp & \text{Here $c$ means ``neither $\ldots$ nor $\ldots$''.} \\[2ex]
  p \lor q & \neg cpq & \text{Ditto}. \\[1ex]
  p \land q & \neg (\neg p \lor \neg q) \\[2ex]
  \exists x. p & \neg \, c \lambda x. p & \text{Here $c$ means ``no $\ldots$ exists''.} \\[1ex]
  \forall x. p & \neg \exists x. \neg p \\[2ex]
  s \neq t & (\lambda x y. \exists z. zx \land \neg zy) st \\[1ex]
  s = t & \neg (s \neq t) \\[2ex]
  p \rightarrow q & \neg p \lor q \\[1ex]
  p \leftrightarrow q & (p \rightarrow q) \land (q \rightarrow p)
\end{array}
\end{displaymath}
The operator priority is as follows from high to low: $\neg$ $=$
$\land$ $\lor$ $\rightarrow$ $\leftrightarrow$ $\lambda$.
\\[1ex]
Variable-binding operators, like $\exists$ and $\forall$, have the
same priority as $\lambda$.
\\[1ex]
Other operators, like $\neq$, have the same priority as $=$ (or $\neg$
if unary).
\\[1ex]
$\lor$ and $\land$ are left associative and $\rightarrow$ and
$\leftrightarrow$ are right associative.
\\[1ex]
$\doteq$ and $\not\doteq$ are similar to $=$ and $\neq$, respectively,
but always has type $\imath \imath o$.
\end{center}
\vspace{0.5ex}
\end{minipage}
\end{center}
\end{boxedminipage}
\end{center}

\begin{center}
\begin{boxedminipage}{\textwidth}
\begin{center}
\begin{minipage}{.96\textwidth}
\vspace{2.5ex}
\begin{center}
$\Theta$, $\Gamma$ and $\Delta$ range over possible empty formula sequences
\\[2ex]
The sequent calculus $\Gamma \,\vdash\, \Delta$ has sequences on both sides in the rules:
\medskip
\begin{eqnarray}
p \,\vdash\, q \quad \text{if $p \sim q$ or $p \succ q$}
\eqname{S}
\end{eqnarray}
\begin{eqnarray}
\prooftree
\Gamma \,\vdash\, \Delta
\Justifies
p,\Gamma \,\vdash\, \Delta \qquad \Gamma \,\vdash\, \Delta,p
\eqname{T}
\endprooftree
\end{eqnarray}
\begin{eqnarray}
\prooftree
\Theta,p,q,\Gamma \,\vdash\, \Delta
\justifies
\Theta,q,p,\Gamma \,\vdash\, \Delta
\endprooftree
\qquad
\prooftree
\Gamma \,\vdash\, \Delta,p,q,\Theta
\justifies
\Gamma \,\vdash\, \Delta,q,p,\Theta
\endprooftree
\eqname{E}
\end{eqnarray}
\begin{eqnarray}
\prooftree
p,p,\Gamma \,\vdash\, \Delta
\justifies
p,\Gamma \,\vdash\, \Delta
\endprooftree
\qquad
\prooftree
\Gamma \,\vdash\, \Delta,p,p
\justifies
\Gamma \,\vdash\, \Delta,p
\endprooftree
\eqname{C}
\end{eqnarray}
\begin{eqnarray}
\prooftree
\Gamma \,\vdash\, \Delta,p,q
\justifies
cpq,\Gamma \,\vdash\, \Delta
\endprooftree
\qquad
\vdash\, p,q,cpq
\eqname{P}
\end{eqnarray}
\begin{eqnarray}
cp,pt \,\vdash\,
\qquad
\prooftree
px,\Gamma \,\vdash\, \Delta
\justifies
\Gamma \,\vdash\, \Delta,cp
\using
~\text{$x$ not free in $p$, $\Gamma$ or $\Delta$}
\eqname{Q}
\endprooftree
\end{eqnarray}
\begin{eqnarray}
\vdash\, p = q \leftrightarrow p \doteq q
\eqname{N}
\end{eqnarray}
\\[1ex]
The left and right rules are placed next to each other.
\\[1ex]
Double lines indicate that the rule works in both directions.
\\[1ex]
Multiple conclusions indicate multiple rules each with a single
conclusion.
\end{center}
\vspace{0.5ex}
\end{minipage}
\end{center}
\end{boxedminipage}
\end{center}

\begin{center}
\begin{boxedminipage}{\textwidth}
\begin{center}
\begin{minipage}{.96\textwidth}
\vspace{-0.5ex}
\begin{center}
\begin{displaymath}
\begin{array}{r@{~\,\coloneqq\,~}l@{\hspace{2em}}l}
\top & \exists x. x & \text{Truth} \\[1ex]
\bot & \neg \top & \text{Falsity} \\[2ex]
s \equiv t & (\lambda x y. \forall z_1 \cdots z_n. x z_1 \cdots z_n \leftrightarrow y z_1 \cdots z_n) st & \text{Equivalence} \\[1ex]
s \not\equiv t & \neg (s \equiv t) \\[2ex]
t' & (\lambda x y. x \doteq y) t & \text{Successor} \\[1ex]
0 & \lambda x. x \not\doteq x & \text{Zero} \\[1ex]
1 & 0' \\[1ex]
2 & 1' \\[1ex]
3 & 2' \quad \ldots
\end{array}
\end{displaymath}
\end{center}
\vspace{-0.5ex}
\end{minipage}
\end{center}
\end{boxedminipage}
\end{center}

\end{document}